\documentclass[aps,prl,twocolumn,showpacs,superscriptaddress]{revtex4}
\usepackage[english]{babel}
\usepackage{graphicx,color}

\usepackage{color}
\definecolor{darkgreen}{rgb}{0,0.5,0}
\definecolor{blue}{rgb}{0,0,0.8}
\definecolor{lightblue}{rgb}{0.93,0.96,1}
\definecolor{darkblue}{rgb}{0.,0.,0.6}
\usepackage[colorlinks,linkcolor=darkgreen,citecolor=darkblue,urlcolor=blue]{hyperref}

\def\reff#1{(\ref{#1})}

\begin{document}

\title{Pattern formation in colloidal explosions\footnote{ Paper published in:\\[2mm]  \href{http://iopscience.iop.org/0295-5075/94/4/48008}{Europhys. Lett. {\bf 94}, 48008 (2011)} \\ DOI: 10.1209/0295-5075/94/48008}  }

\author{Arthur V. Straube}
\affiliation{Department of Physics, Humboldt University of Berlin, Newtonstr. 15, D-12489 Berlin, Germany}
\affiliation{Rudolf Peierls Centre for Theoretical Physics, University of Oxford, 1 Keble Road, Oxford OX1 3NP, United Kingdom}

\author{Ard A. Louis}
\affiliation{Rudolf Peierls Centre for Theoretical Physics, University of Oxford, 1 Keble Road, Oxford OX1 3NP, United Kingdom}

\author{J{\"o}rg Baumgartl}
\affiliation{2. Physikalisches Institut, Universit{\"a}t Stuttgart, Pfaffenwaldring 57, D-70569 Stuttgart, Germany}

\author{Clemens Bechinger}
\affiliation{2. Physikalisches Institut, Universit{\"a}t Stuttgart, Pfaffenwaldring 57, D-70569 Stuttgart, Germany}

\author{Roel P. A. Dullens}
\affiliation{Physical and Theoretical Chemistry Laboratory, Department of Chemistry, University of Oxford, South Parks Road, Oxford OX1 3QZ, United Kingdom}

\begin{abstract}
We study the non-equilibrium pattern formation that emerges when magnetically repelling colloids, trapped by optical tweezers, are abruptly released, forming colloidal explosions. For multiple colloids in a single trap we observe a pattern of expanding concentric rings.  For colloids individually trapped in a line, we observe explosions with a  zigzag pattern that persists even when magnetic interactions are much weaker than those that break the linear symmetry in equilibrium.   Theory and computer simulations quantitatively describe these phenomena both in and out of equilibrium. An analysis of the  mode spectrum allows us to accurately quantify the non-harmonic nature of the optical traps. Colloidal explosions provide a new way to generate well-characterized non-equilibrium behaviour in colloidal systems.
\end{abstract}





\maketitle

\textbf{Introduction.} -- Pattern formation is an important and widespread phenomenon in the natural world and provides striking examples of order produced by non-equilibrium processes~\cite{Cross-Hohenberg-93}. For instance, the confinement of charged particles to spatially localized traps can lead to one-dimensional Wigner crystals or the spontaneous formation of zigzag or helical particle patterns.   Such confined particle geometries are often used for ion storage and quantum-computing experiments~\cite{Birkl-etal-92, GarciaMata-etal-07}.  But when the confinement is too weak, the particles can escape from the traps.  Often, particle trajectories produced from such ``explosions'' proceed in an incoherent manner, but in the case of long-ranged interactions, more interesting spatio-temporal patterns may form. In this letter, we investigate the dynamical behaviour of strongly interacting superparamagnetic colloidal particles that have first been confined with an optical laser trap. When the trap is abruptly turned off, the long-ranged repulsive inter-colloidal interactions generate a ``colloidal explosion'' with characteristic non-equilibrium patterns that depend on the initial conditions and on the strength of the interactions.

\begin{figure*}[!htb]
\includegraphics*[width=16.5 cm]{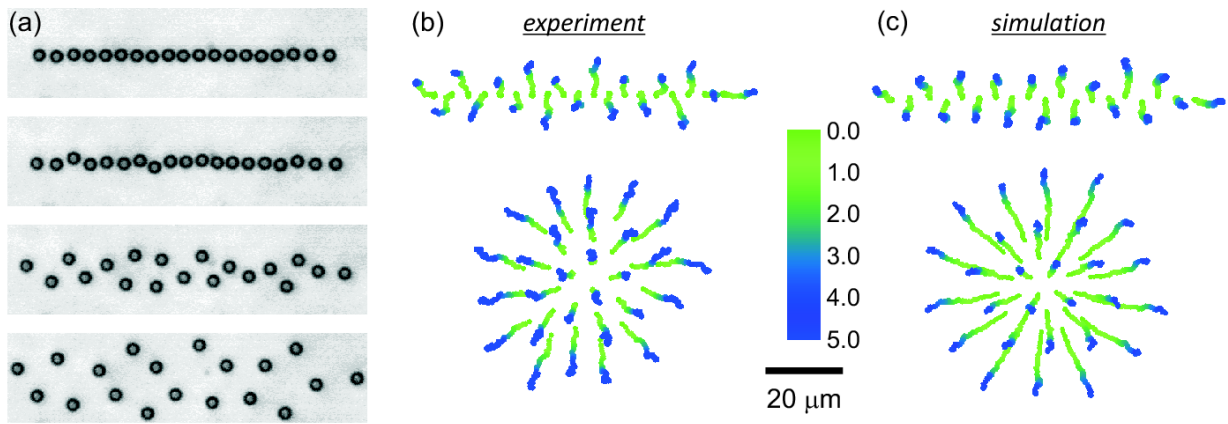}
\centering
\caption{(Color online) (a) Microscopy images ($92 \times 23\,\mu$m$^2$) showing an exploding 1d array of $N=19$ magnetic particles with a spacing $L = 4 \,\mu{\rm m}$ upon removing the optical traps at $t=0 \, s$ at an external magnetic field $B=1.90\,{\rm mT}$ and stiffness $k_0=0.37 \pm 0.01 \,{\rm pN/\mu m}$. From top to bottom: $t < 0 \, s$, $0.2\, s$, $1 \,s$, and $5 \, s$. (b) The experimental particle trajectories of the exploding chain (top) and for an exploding 2d colloidal disc (bottom) of $N=28$ particles at a magnetic field of $B=2.25\,{\rm mT}$. The color code indicates the time in seconds. (c) Brownian dynamics trajectories for an exploding chain (top) and for a colloidal disc compare well to the experiments.
}
\label{explosion}
\end{figure*}

Colloidal systems offer the unique ability to simultaneously visualize and carefully control the non-equilibrium behaviour using external fields, which allows for detailed comparisons between experiments and theory~\cite{Loewen-01}. Recent examples include non-linear instabilities in sedimenting suspensions~\cite{Wysocki-etal-09}, dynamic lane formation in oppositely charged particles under electric fields~\cite{Leunissen-etal-05}, driven dislocation nucleation~\cite{Schall-etal-06} and stochastic resonance~\cite{Babic-etal-04}. Here we exploit the ability to exquisitely tune inter-colloidal magnetic interactions through the application of an external magnetic field~\cite{Bubeck-etal-99} while simultaneously placing colloids into a well-defined initial configuration using optical tweezers. The single-particle trajectories are directly monitored by video-microscopy \cite{Keim-etal-04}. To explore the physics of colloidal explosions we study two basic geometries, namely the one-dimensional (1d) and two-dimensional (2d) configurations shown in Fig.~\ref{explosion}.   Even thought these geometries are fairly simple, they exhibit non-trivial non-equilibrium behaviour. This study opens up the possibility of exploring many other geometries and explosion patterns.

\vspace{3mm}\textbf{Experimental and simulation details.} -- We use superparamagnetic latex spheres of radius $a=1.35$~$\mu$m (Dynabeads, Invitrogen) in a water solvent, contained in a 200 $\mu$m thick quartz glass sample cell. The gravitational length of the particles is much smaller than their diameter so after sedimentation this is effectively a 2d system. The application of a perpendicular magnetic field induces long-range repulsive magnetic interactions between the colloids of the form
\begin{equation}
V^{mag}_{mn}=\frac{\mu_0}{4 \pi}\frac{\chi^2 B^2}{r_{mn}^3},
\label{repulsive-pot}
\end{equation}
where \noindent  $\mu_0$ is the permeability of free space, $B$ is the magnetic field, and  $r_{mn}=|{\mathbf r}_{m}-{\mathbf r}_{n}|$ is the distance between particles $n$ and $m$, and the colloids have a magnetic susceptibility of $\chi=3.95\times$10$^{-12}{\rm\;A\;m}^2$/T~\cite{Blickle-etal-05}. The particle trajectories are obtained using video-microscopy \cite{Keim-etal-04, Crocker-Grier-96}.

The colloids are trapped using acousto-optical-deflection controlled optical laser tweezers (see e.g. \cite{Babic-etal-04}). The trapping potential for particle $n$ is modeled as a
\begin{equation}
V^{trap}_n=V_0\left[1-\exp\left(-\frac{k_0}{2V_0} \delta r_n^2\right)\right], \label{trap-pot}
\end{equation}
\noindent which takes into account, in a generic way, the fact that the trap has a finite range of attraction. Here,  ${\mathbf r}_n$ is the particle position, ${\mathbf R}_n=nL\hat{\mathbf x}$ is the trap position, $\delta {\mathbf r}_n={\mathbf r}_n-{\mathbf R}_n$ and $V_0$ is the depth of the potential well. The softness of the trap is characterized by  the dimensionless parameter
%
\begin{equation}
\alpha = \frac{k_0 L^2}{V_0}. \label{alpha-def} \nonumber
\end{equation}
When $\alpha = 0$ the trapping is purely quadratic, $V^{trap}_n=k_0 \delta r_n^2/2$, for all $\delta r_n$,  whereas for $\alpha > 0$ the trap is quadratic for small $\delta r_n$, but  becomes increasingly non-harmonic at larger $\delta r_n$ and has a finite height $V_0$ above which the particle can escape from the trap.
  It is important to note that  $k_0$ and $V_0$ can be changed in the experiment but that their ratio remains virtually constant and therefore characterizes the optical trap~\cite{Bechinger-etal-01, Dobnikar-etal-04}.

The maximum particle velocities observed in the experiments were of order  $2 \; \mu{\rm m/s}$ in the explosion so that the Reynolds number  ${\rm Re}\sim 10^{-5}$ is small. Simple overdamped Brownian Dynamics (BD) simulations with a single fit parameter to the field $B$ in Eq.~(\ref{repulsive-pot}) were found to closely mimic the experimental trajectories as demonstrated in Fig.~\ref{explosion}(b) and (c). Combining the repulsive and trapping potentials, Eqs.~\reff{repulsive-pot} and \reff{trap-pot}, at finite temperature $T$ leads to the following dimensionless equation of motion
\begin{equation}
\dot{\mathbf r}_n  =  3b^2\sum_{m\ne n}\frac{{\mathbf r}_{mn}}{r_{mn}^5}-\delta {\mathbf r}_n\exp\left(-\frac{\alpha}{2}\delta r_n^2\right)+\sqrt{2\epsilon}\,{\mathbf s}_n. \label{BD-eq-dimless}
\end{equation}
\noindent Here and afterwards, the length, the energy, and the time are expressed in terms of of $L$, $k_0 L^2$, and $6\pi\eta a/k_0$ ($\eta$ is the dynamic viscosity of the solvent), respectively. The stochastic force ${\mathbf s}(t)$ obeys the properties $\left<{\mathbf s}_n(t)\right>=0$ and $\left<{\mathbf s}_n(t){\mathbf s}_{n'}(t')\right>=\delta_{nn'}\delta(t-t')$ with the first $\delta$ being Kronecker's delta function and the second being Dirac's. Apart from the parameter characterizing the softness of the trapping potential, $\alpha$, the dynamics is governed by dimensionless parameters
%
\begin{equation}
b^2=\frac{\mu_0\chi^2 B^2}{4\pi k_0 L^5}, \qquad \epsilon = \frac{k_B T}{k_0 L^2}. \label{b2-apsilon-def} \nonumber
\end{equation}
These two additional parameters describe respectively the intensity of the magnetic field and the intensity of thermal motion. The fact that we work at a given temperature $T$, and at certain values of $L$ and $k_0$, fixes the value of $\epsilon$. To integrate Eq.~\reff{BD-eq-dimless}, we apply a standard algorithm \cite{Ermak-McCammon-78}.

\vspace{3mm}\textbf{Explosion of a disc.} -- We first describe the 2d geometry dealing with explosions from a disc composed of a variable number of colloids, as shown for an example with $N=28$ particles in Fig.~\ref{explosion}(b). To create such a disc configuration in our simulations, all the colloids are captured by a wide single trap, ${\mathbf R}_n=(0,0)$. However, as we argue below, our experimental optical trap is characterized by the value $\alpha \approx 30$, implying that the effective entrapment range is such that the trap cannot easily hold more than one colloid at a time. In the experiments, we therefore gathered all the colloids together into a disc, which was easily done with a single trap in the absence of the magnetic field. As soon as the colloids have been gathered in the disc, the trap was switched off and, at the same time, the magnetic field was quickly switched on. Typically, this procedure is done within the time $\Delta\tau \le 1 \,{\rm s}$. As the characteristic diffusion time for the particles used in experiments is $\tau_D \sim a^2/D \approx 10 \,{\rm s}$ (where $D$ is the particle diffusivity), the condition $\Delta\tau < \tau_D$ ensures that the gathered colloids have not significantly diffused away from their initial configuration.

We observe a clear pattern of concentric rings as the colloids move to minimize the magnetic repulsion between them (see Fig.~\ref{explosion}(b) and (c)). The velocities $v$ of particles are maximal in the beginning of the explosion and decay as the particles move apart. As the repulsive interactions are long ranged, in an unbounded domain without thermal fluctuations  this process never stops and the particles move out to infinity. Nevertheless, at finite temperature a  characteristic explosion time $\tau_e$ can be defined as the time to reach a regime of motion with a small enough velocity such that the P\'eclet number ${\rm Pe}=av/D\sim 1$. At smaller velocities, diffusion starts to dominate.

\begin{figure}[!tb]
\begin{center}
\centering\includegraphics[width=0.45\textwidth]{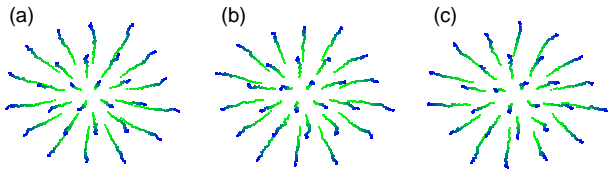}
\end{center}
\caption{(Color online) Particle trajectories for explosions from two-dimensional discs of $N=28$ particles with and without delay. The values of parameters $b=0.274$ and $\alpha=0$ are chosen to provide the best fit to the experiment. (a): no delay, $\Delta\tau=0$; (b): delay time $\Delta\tau \approx \tau_D/10$; (c): delay time $\Delta\tau \approx \tau_D/5$.
} \label{fig:disc-delays}
\end{figure}

In Fig.~\ref{fig:disc-delays} we illustrate simulated trajectories for explosions of a disc of $N=28$ particles for different delays $\Delta\tau$, in which the explosion time $\tau_e\approx \tau_D/6$. Note that the trapped colloids can come very close to each other. To ensure that the colloids do not overlap in simulations, we have additionally included in Eq.~\reff{BD-eq-dimless} steep repulsive interactions of the Weeks-Chandler-Andersen form \cite{Hansen-McDonald-86}. These results confirm that our way of creating a disc in the experiment is practically similar to the use of a single wider trap in the simulations as described above. We emphasize that the results remain robust even for delay times that are longer than in the experiment and comparable with the time of explosion, as, e.g., in Fig.~\ref{fig:disc-delays}(c).

%
%

\begin{figure*}[!htb]
\includegraphics*[width=16 cm]{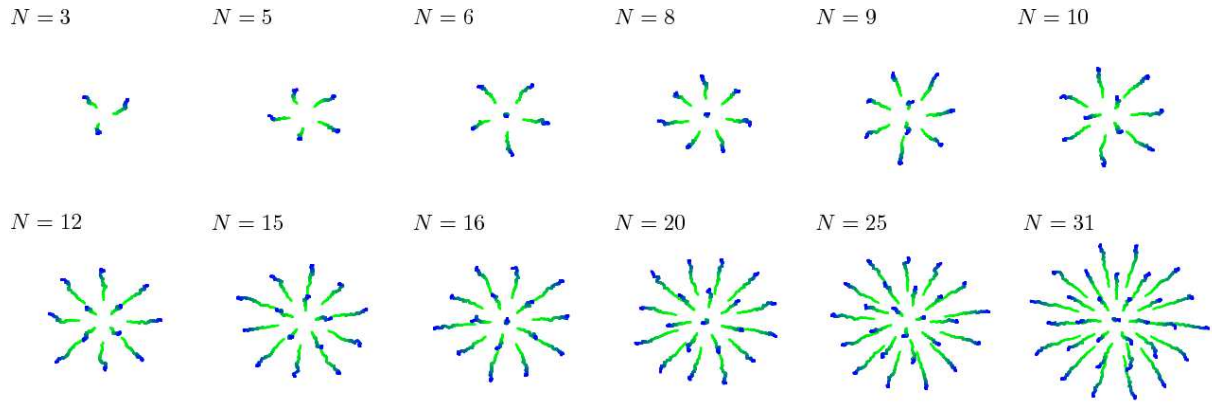}
\centering
\caption{(Color online) Particle trajectories for explosion from two-dimensional discs composed of different number $N$ of colloids. Parameters are $b=0.274$, $\alpha=0$. }
\label{clusters-explosion}
\end{figure*}

In figure~\ref{clusters-explosion} we show simulation trajectories for discs composed of different 
number of colloids.
The initial configuration in the form of a 2d disc can be considered as a pattern of concentric rings that helps explain the non-equilibrium pattern seen in the explosions. As $N$ increases, the number of rings grows, with a single particle remaining near the centre for $N=6$, marking the beginning of a second ring, and then again at $N=16$, marking the beginning of a third ring, and again at $N=31$, marking the beginning of a fourth ring. The pattern formation is caused by the initial shell-like ordering of the particles that is driven by the combination of repulsive magnetic fields and confinement~\cite{Bubeck-etal-99}. The particles in each shell move outwards in the same manner until the repulsive magnetic interactions between the particles are less than $k_B T$ ($k_B$ is Boltzmann's konstant) and diffusion begins to dominate. We note that this colloidal system is reminiscent of   Coulomb explosions induced by strong laser fields stripping off the electrons of molecules and atomic clusters~\cite{Wabnitz-etal-02, Ebeling-Romanovsky-09}.

\vspace{3mm}\textbf{Explosion of a chain.} -- Whereas the broad features of the colloidal disc explosions are fairly straightforward to explain, the behaviour of the 1d configuration is more subtle.  This is not unexpected since 1d chains of interacting particles form some of the richest and most important models in theoretical physics.  Examples include the Frenkel-Kontorova model \cite{Braun-Kivshar-91}, 1d models of Wigner \cite{Schulz-93,Piacente-etal-prb-04}, colloidal~\cite{Polin-etal-06,Leonardo-etal-07} and microfluidic droplets \cite{Beatus-etal-06, Beatus-etal-07} crystals and other closely related systems such as polymer chains \cite{Doi-Edwards-86} and dusty plasmas \cite{Melzer-06}.


The experiment shown in Fig.~\ref{explosion}(a) and (b) is for $N=19$ colloids placed with a lattice spacing of $L=4.00\,\mu{\rm m}$ apart. When the optical traps are removed in the presence of the magnetic field, the particle chain explodes and the evolving pattern shows an almost perfect zigzag symmetry. To analyze the underlying equilibrium behaviour in the presence of a magnetic field, but before the optical traps are switched off, i.e. before the explosion, we begin by treating the deterministic ($\epsilon=0$) case for $N \rightarrow \infty$ colloids. For a strong enough inter-colloidal repulsions, and for strong enough optical traps, we expect that the linear symmetry will be broken, leading to an equilibrium transition from a linear to a quasi-one-dimensional zigzag state as, e.g., in the Frenkel-Kontorova model \cite{Braun-Kivshar-91}, dusty plasmas~\cite{Melzer-06}, quantum wires~\cite{Meyer-etal-07} and 1d Wigner crystals~\cite{Piacente-etal-prb-04}. The zig-zag state can be described by the order parameter $h$, where
 ${\mathbf h}_n/L=n\hat{\mathbf x}+(-1)^n (h/2)\hat{\mathbf y}$  is the equilibrium displacement from the centre of the trap of particle $n$.
 Combining the induced interaction~(\ref{repulsive-pot}) with the laser traps potential~(\ref{trap-pot}) leads to an
 interaction energy (per particle) $U(h)$ that can be written in the dimensionless form
\begin{equation}
U(h) =b^2\sum_{m=1}^{\infty} f_{m}^{3}(h)+\frac{1-\exp(-\alpha h^2/8)}{\alpha} \label{pot-U0},
\end{equation}
\noindent with $f_m(h)=(m^2+p_m h^2)^{-1/2}$ and $p_m=[1-(-1)^m]/2$.
%
The equilibrium displacement $h_{*}$  can be found as a function of $b$ and $\alpha$ by minimizing the potential~(\ref{pot-U0}).
The critical field $b_c$ at which the transition from the line to the zigzag state with $h=h_{*} \ne 0$ occurs  is found to be $b_c = \sqrt{8/[93 \zeta(5)]}\approx 0.288$, with $\zeta(x)=\sum_{m=1}^{\infty}m^{-x}$ the Riemann Zeta function.  Close to the critical point, where $|b-b_c| \ll b_c$ and $h_{*} \ll 1$, we obtain a square-root law
\begin{equation}
h_{*} = \pm \sqrt{\frac{8(b^2-b_c^2)}{b_c^2(\alpha_c-\alpha)}} \label{h_m-small}
\end{equation}
\noindent with $\alpha_c=635 \zeta(7)/[31 \zeta(5)] \approx 19.9$. We note that the nearest neighbor (NN) approximation, where only one term with $m=1$ in relation \reff{pot-U0} is retained, works very well: We obtain $b_c^{NN}=1/\sqrt{12}\approx 0.289$ and $\alpha_c^{NN}=20$.

We note that the existence of the equilibrium zig-zag state is determined by the value of the softness parameter, $\alpha$. For potentials with $\alpha < \alpha_c$, the 1d line state is stable for $b< b_c$, whereas for $b > b_c$ the  zig-zag state is stable and the displacement $h_*$  grows with increasing $b$.  However, for   $0  < \alpha < \alpha_c$, increasing the field further eventually leads to an instability at a larger field $b_{**}$ where the equilibrium zig-zag state is unstable, i.e.\  the barrier to the particles leaving the wells disappears. Within the NN approximation we find that $b_{**} = b_c^{NN} (\alpha_c^{NN}/\alpha)^\frac{5}{4} \exp[-(\alpha_c^{NN}-\alpha)/16]$, although finite temperature means particles can escape at lower fields, $b_{**}^{\prime}=b_{**}\{1-(1/40)\exp[(\alpha_c-\alpha)/12](30 k_BT/V_0)^{2/3}\}$. For softer potentials with $\alpha > \alpha_c$, there is no transition to an {\em equilibrium} zigzag state. In this case, the particles experience a colloidal explosion and leave their wells before an equilibrium zigzag state can occur. In the absence of thermal fluctuations, increasing the field would yield an explosion of the 1d line state at $b=b_c$. At finite temperature, particles can escape at smaller fields, $b_c^{\prime} \approx b_c[1-\sqrt{(1-\alpha_c/\alpha)k_B T/2V_0}]$.

\begin{figure}[!t]
\centering \includegraphics*[width=8.0 cm]{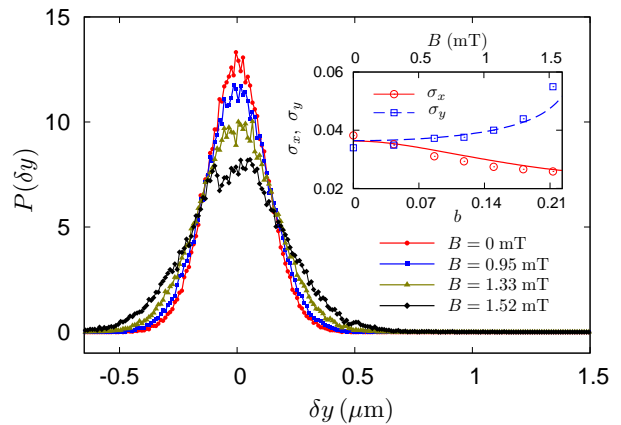}
\caption{(Color online) Probability density distribution in the transversal direction,  $P(\delta y)$, for increasing magnetic field $B$. The inset shows respectively the standard deviations $\sigma_x$ (solid line) and $\sigma_y$ (dashed line) of $P(\delta x)$ and $P(\delta y)$ as functions of magnetic field. Points represent experimentally measured data, the lines provide the fit obtained via BD simulations with $\alpha=30$.}
\label{pyplot}
\end{figure}

 The question then arises whether for the experiments set-up used in Fig.~\ref{explosion}, $\alpha$ is small enough to allow an equilibrium zig-zag transition. To test this we measured the probability density functions for displacements along ($P(\delta x)$) and perpendicular to ($P(\delta y)$) the trapped chain for different magnetic fields. As can be seen in  Fig.~\ref{pyplot}, $P(\delta y)$ widens for increasing magnetic fields, but always retains its single-peaked structure. This confirms that the system is \emph{not} in an equilibrium zigzag state, despite the zigzag symmetry in the explosion pattern originating from these states. The inset of Fig.~\ref{pyplot} presents respectively the standard deviations $\sigma_x$ and $\sigma_y$ for distributions $P(\delta x)$ and $P(\delta y)$ as functions of the magnetic field.
 The observation that $\sigma_x$ decreases and $\sigma_y$ increases upon increasing the magnetic field points to respectively hardening of the longitudinal and  softening of transverse normal modes~\cite{Piacente-etal-prb-04}.

We also measured the phonon dispersion relations by tracking the particle displacements ${\mathbf u}({\mathbf h}_n)$ from their equilibrium positions ${\mathbf h}_n$. The Fourier transforms ${\mathbf u}(q)$ of the displacement vectors are directly related to the dynamical matrix \ $D_{\mu\nu}(q)$~\cite{Keim-etal-04}:
\begin{equation}
\left < u_\mu^*(q) u_{\nu}(q) \right > = k_BT \,D_{\mu \nu}^{-1}(q),
\label{dynmatr-rel}
\end{equation}
\noindent where the average is over all independent configurations.  The eigenvalues of $D_{\mu\nu}(q)$ yield the normal mode spring constants.
\begin{figure}[!t]
\centering \includegraphics*[width=8.0 cm]{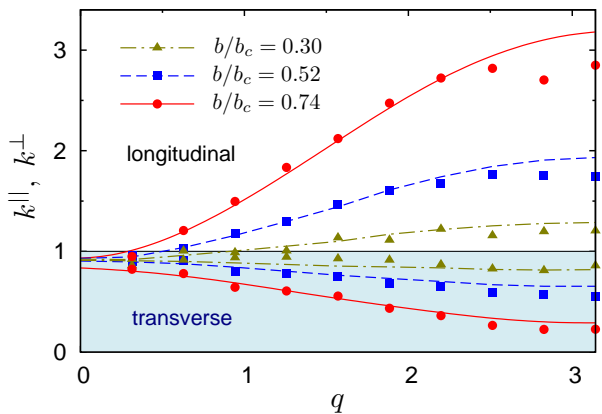}
\caption{(Color online) The longitudinal and transverse normal mode spring constants $k^{||,\perp}$ for different magnetic fields as a function of the wave number $q$. Points are experimental data. Lines are from BD simulations with $\alpha=30$.}
\label{phononQ}
\end{figure}
Figure \ref{phononQ} shows the phonon-dispersion relations as a function of $q$ for different values of the magnetic field and the hardening and softening of the longitudinal and transverse modes respectively is indeed evident.

We analyzed these results further using  BD simulations.  Given the spring constant, the interparticle distance $L$ and the temperature, there are only two parameters left to determine, the ratio between $B$ and $b$ and the potential softness parameter $\alpha$.  A fit to the data in both Fig.~\ref{pyplot} and Fig.~\ref{phononQ} yields a single proportionality constant $B/b$ and also fixes $\alpha = 30 \pm 5$.   There is good  {\em quantitative} agreement with the data in both figures.  The values of $b < b_c$, suggest that we are below the critical field, and furthermore the fact that $\alpha > \alpha_c$ suggests that for this experimental system no equilibrium zigzag state is possible.

  We note that varying $\alpha$ affects the transverse and longitudinal modes differently.  For $\alpha=0$, $k^{||,\perp}(q=0)=1$ ($b<b_c$), whereas for $\alpha \ne 0$ we obtain at $q=0$: $k^{||}\ne k^{\perp}\ne 1$, in agreement with experiment.  Note that $k^{||,\perp}$ are in units of $k_0$. These deviations can be traced back to the softness of the trapping potentials.  At finite temperature, the particles explore the  non-harmonicity of the potential, leading to what appears to be a weaker effective spring force. As before, the accurate NN approximation helps illuminate these results.
The  normal modes  correspond to ${\mathbf u}({\mathbf h}_n)\propto\hat{\mathbf n} \exp(-\lambda t +i n q)$, where $\lambda$ is the decay rate, $q \in [0,\pi]$ is the wave number, and $\hat{\mathbf n}=\hat{\mathbf x}$ or $\hat{\mathbf y}$. For the line state we obtain, in dimensionless form,
\begin{eqnarray}
k_{NN}^{||}&=&1+4\left(\frac{b}{b_c^{NN}}\right)^2\sin^2\left(\frac{q}{2}\right), \label{kNN-long} \\
k_{NN}^{\perp}&=&1-\left(\frac{b}{b_c^{NN}}\right)^2\sin^2\left(\frac{q}{2}\right) \label{kNN-perp},
\end{eqnarray}
%
\noindent from which we draw three important conclusions.

First, this result shows hardening and softening of the spring constants with the field in respectively the longitudinal and transverse directions. Since  $k^{||,\perp} \propto \sigma_{x,y}^{-1}$, this is consistent with the inset of Fig.~\ref{pyplot}. Note that the dependencies of $k^{||,\perp}$ on $q$ at a given $b$ provided by Eqs.~\reff{kNN-long} and \reff{kNN-perp} are in qualitative agreement with those obtained via relation \reff{dynmatr-rel} and presented in Fig.~\ref{phononQ}. Second, Eqs.~\reff{kNN-long} and \reff{kNN-perp} show that as $b$ goes beyond $b_c$, the mode that first becomes unstable is the zigzag mode, $q=\pi$.
Third, in a system with a finite number of particles the spectrum of decay rates is discrete. There are other eigenvalues present as well
(cf. markers in Fig.~\ref{phononQ}), so that the zigzag mode is better separated from other eigenmodes in chains with smaller number of beads, $N$. Hence, while the perfect zigzag pattern is the most probable in explosion of shorter chains, the probability of imperfections in this pattern increases with the chain length. We indeed observed this effect in simulations of longer chains.

\begin{figure}[!tb]
\centering\includegraphics[width=0.45\textwidth]{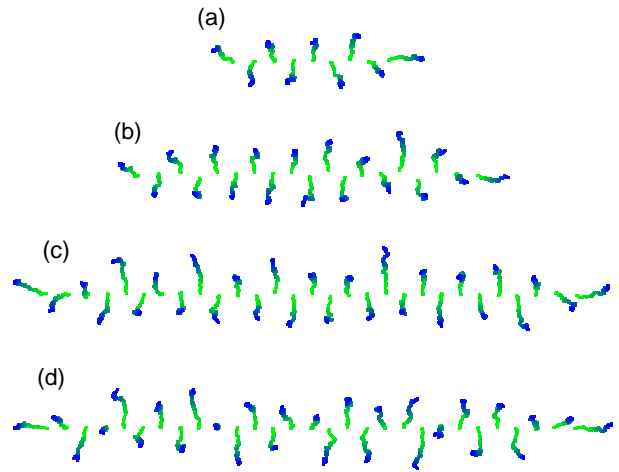}
\caption{(Color online) Particle trajectories showing zig-zag explosions for chains with different number of colloids: $N=9$ (a), $N=19$ (b), and $N=29$ (c). d) A pattern with defects for $N=29$. Parameters are: $b=0.8 b_c$, $\alpha=30$.
} \label{fig:chain-diff-N}
\end{figure}

In Fig.~\ref{fig:chain-diff-N} we show explosion patterns for 1d chains of different lengths. We fix the magnetic field to a subcritical value $b=0.8 b_c$ and the softness parameter to $\alpha=30$ (as in the experiments) which guarantees that the explosion occurs from the equilibrium line state.  As can be seen from the trajectories, similar explosion patterns with the zig-zag symmetry can be found in shorter and longer chains. However, for the longer chains, explosion patterns with defects also occur, see Fig.~\ref{fig:chain-diff-N}(d). The simplest defect corresponds to a pair of neighbouring colloids shooting out in the same direction. Defects appear because phonon modes other than the zigzag mode play an increasingly important role with increasing $N$.
%
%
%

For systems with $\alpha < \alpha_c$  an equilibrium zigzag state can always be induced when $b_{**}' >b > b_c$. In contrast to systems without an equilibrium zigzag state, we observe perfect zigzag symmetry in the explosions for all $N$  we investigated.  The different behaviour results because in this case the non-equilibrium pattern simply reflects the broken symmetry in equilibrium.

 The fact that a single fit to $B/b$ and $\alpha$ provides an accurate fit to all the data from Figs.~\ref{pyplot} and~\ref{phononQ}, combined with the single-peaked probability distributions, strongly suggests that the explosions observed in Fig.~\ref{explosion} are for the situation where there is no equilibrium zigzag pattern. The zigzag pattern we observe is entirely due to a non-equilibrium pattern formation effect that shadows the hidden equilibrium symmetry breaking.

Finally, we also considered the role of hydrodynamic interactions (HI) by using 3d Stokesian dynamics.  We find very similar behaviour to that seen with BD simulations that neglect the HI.
The real HI will be modulated by the surface, but given the small overall effect of HI, we argue that explicitly including them is not important  for the colloidal explosions we studied.

\vspace{4mm}\textbf{Conclusions.} -- In summary, we have exploited the ability to carefully control and characterize a colloidal model system to generate non-equilibrium pattern formation in colloidal explosions.  For a 2d initial geometry, we observe patterns with expanding concentric rings that can be explained by the initial shell-like ordering.  For the 1d geometry, we observe explosions with a zigzag pattern at fields strengths well below those that would break the linear symmetry in equilibrium.   Furthermore, a quantitative comparison to phonon dispersion relationships allows us to characterize the non-harmonic nature of the traps.

More generally, we have introduced ``colloidal explosions,'' a new way to generate well-characterised non-equilibrium behaviour in colloidal systems.  It should be possible to create such explosions with a range of different kinds of confining potentials and repulsive inter-particle interactions.   This methodology can be applied to a wide variety of other geometries, leading to potential applications, for example, in microfluidics.

\vspace{4mm}\textbf{Acknowledgements.} -- We are grateful to P.~Chaikin, L.~Schimansky-Geier, M.~Zaks, S. Shklyaev and A.~Pikovsky for fruitful discussions. AS was supported by German Science Foundation (DFG Project No. STR 1021/1-2) and by HPC-EUROPA2 (Project No. 228398). AAL acknowledges the Royal Society for financial support. RPAD acknowledges the Alexander von Humboldt Foundation for financial support.


\begin{thebibliography}{10}

\bibitem{Cross-Hohenberg-93}
M.~C.~Cross and P.~C.~Hohenberg,
\newblock Rev. Mod. Phys. {\bf 65}, 851 (1993).

\bibitem{Birkl-etal-92}
G.~Birkl, S.~Kassner, and H.~Walther,
\newblock Nature (London) {\bf 357}, 310 (1992).

\bibitem{GarciaMata-etal-07}
I.~Garc{\'i}a-Mata, O.~V.~Zhirov, and D.~L.~Shepelyansky,
\newblock Eur. Phys. J. D {\bf 41}, 325 (2007).

\bibitem{Loewen-01}
H. L\"{o}wen,
\newblock J. Phys. Condens. Matter {\bf 13} R415 (2001).

\bibitem{Wysocki-etal-09}
A. Wysocki {\em et al.},
\newblock Soft Matter {\bf 5}, 1340 (2009);
J.~T. Padding and A. A. Louis, Phys. Rev. E \textbf{77}, 011402 (2008).


\bibitem{Leunissen-etal-05}
M.~E.~Leunissen {\em et al.},
\newblock Nature (London) {\bf 437}, 235 (2005).

\bibitem{Schall-etal-06}
P.~Schall, I.~Cohen, D.~A.~Weitz, and F.~Spaepen,
\newblock Nature (London) {\bf 440}, 319 (2006).

\bibitem{Babic-etal-04}
D.~Babi\v{c}, C.~Schmitt, I.~Poberaj, and C.~Bechinger,
\newblock Europhys. Lett. {\bf 67}, 158 (2004).

\bibitem{Bubeck-etal-99}
R.~Bubeck, C.~Bechinger, S.~Neser, and P.~Leiderer,
\newblock Phys. Rev. Lett. {\bf 82}, 3364 (1999).

\bibitem{Keim-etal-04}
P.~Keim, G.~Maret, U.~Herz, and H.~H. von Gr{\"u}nberg,
\newblock Phys. Rev. Lett. {\bf 92}, 215504 (2004).

\bibitem{Blickle-etal-05}
V.~Blickle, D.~Babi\v{c}, and C.~Bechinger,
\newblock Appl. Phys. Lett. {\bf 87}, 101102 (2005).

\bibitem{Crocker-Grier-96}
J.~C.~Crocker and D.~G.~Grier,
\newblock J. Colloid Interface Sci. {\bf 179}, 298 (1996).

\bibitem{Bechinger-etal-01}
C.~Bechinger, M.~Brunner, and P.~Leiderer,
\newblock Phys. Rev. Lett. {\bf 86}, 930 (2001).

\bibitem{Dobnikar-etal-04}
J.~Dobnikar, M.~Brunner, H.-H.~von Gr{\"u}nberg, and C.~Bechinger,
\newblock Phys. Rev. E {\bf 69}, 031402 (2004).

\bibitem{Ermak-McCammon-78}
D.~L.~Ermak and J.~A.~McCammon,
\newblock J. Chem. Phys. {\bf 69}, 1352 (1978).

\bibitem{Hansen-McDonald-86}
J.~P.~Hansen and I.~R.~McDonald,
\textit{Theory of Simple Liquids, 2nd ed.} (Academic Press, London, 1986).

\bibitem{Wabnitz-etal-02}
H.~Wabnitz {\em et al.},
\newblock Nature (London) {\bf 420}, 482 (2002).

\bibitem{Ebeling-Romanovsky-09}
W.~Ebeling and M.~Yu. Romanovsky,
\newblock Contrib. Plasma Phys. {\bf 49}, 477 (2009).

\bibitem{Braun-Kivshar-91}
O.~M.~Braun and Y.~S.~Kivshar,
\newblock Phys. Rev. B {\bf 44}, 7694 (1991).

\bibitem{Schulz-93}
H.~J.~Schulz,
\newblock Phys. Rev. Lett. {\bf 71}, 1864 (1993).

\bibitem{Piacente-etal-prb-04}
G.~Piacente, I.~V.~Schweigert, J.~J.~Betouras, and F.~M.~Peeters,
\newblock Phys. Rev. B {\bf 69}, 045324 (2004).

\bibitem{Polin-etal-06}
M.~Polin, D.G.~Grier, and S.~R.~Quake,
\newblock Phys. Rev. Lett. {\bf 96}, 088101 (2006).

\bibitem{Leonardo-etal-07}
R.~Di~Leonardo {\em et al}.,
\newblock Phys. Rev. E {\bf 76}, 061402 (2007).

\bibitem{Beatus-etal-06}
T.~Beatus, T.~Tlusty, and R.~Bar-Ziv,
\newblock Nature Phys. {\bf 2}, 743 (2006).

\bibitem{Beatus-etal-07}
T.~Beatus, R.~Bar-Ziv, and T.~Tlusty,
\newblock Phys. Rev. Lett. {\bf 99}, 124502 (2007).

\bibitem{Doi-Edwards-86}
M.~Doi and S.~Edwards,
\textit{The Theory of Polymer Dynamics} (Oxford University Press, Oxford, 1986).

\bibitem{Melzer-06}
A.~Melzer,
\newblock Phys. Rev. E {\bf 73}, 056404 (2006).

\bibitem{Meyer-etal-07}
J.~S.~Meyer, K.~A.~Matveev, and A.~I.~Larkin,
\newblock Phys. Rev. Lett. {\bf 98}, 126404 (2007).

\end{thebibliography}
\end{document}